\documentclass[useAMS,usenatbib]{mn2e}
\usepackage{graphicx}
\usepackage{textcomp}
\usepackage{epstopdf}
\usepackage{gensymb}

\title[Discovery of an optical and X-ray synchrotron jet in NGC 7385]{Discovery of an optical and X-ray synchrotron jet in NGC 7385}

\author[J. Rawes, D. M. Worrall and M. Birkinshaw]{J. Rawes, \thanks{E-mail:
jr9420@bristol.ac.uk} D. M. Worrall and M. Birkinshaw\\
H.H. Wills Physics Laboratory, University of Bristol, Tyndall Avenue, Bristol BS8 1TL, UK}

\begin{document}

\date{Accepted ... Received ...; in original form ...}

\pagerange{\pageref{firstpage}--\pageref{lastpage}} \pubyear{2014}

\maketitle

\label{firstpage}

\begin{abstract}
 We report the discovery of optical and X-ray synchrotron emission from the brighter radio jet in galaxy NGC 7385 using data from HST and \textit{Chandra}. The jet has a projected length of 5 kpc and a similar morphology to other known optical jets in low-power radio galaxies.  We also report a strong jet-cloud interaction which appears to be deflecting the counter-jet and causing a reversal in its direction.
\end{abstract}


\section{Introduction}

Relativistic outflows of plasma from the central regions of active galactic nuclei (AGN) are often observed as jets extending away from the nuclei of their parent galaxies. At large distances these may develop into large-scale `plumes' or radio lobes. Jets and lobes play host to some of the most energetic particles in the Universe. Despite the large number of radio jets that have X-ray counterparts, relatively few optical counterparts have been detected. \footnote{See http://astro.fit.edu/jets and references therein}

Radio jets have been intensely studied with high-resolution interferometry. Their radio emission is synchrotron radiation, which is often highly polarised (e.g. Bridle \& Perley 1984). Transport of energy and momentum by jets is important in the context of structure formation.

Optical synchrotron jets are of particular interest because the energy loss times for high-energy electrons are typically short (shorter still for X-rays), of the order of 100 years. X-ray and optical synchrotron features localise the sites of particle acceleration, thus studies of optical jets can probe jet interiors and constrain energy transfer processes in radio galaxies. Because of this potential importance to radio-source physics, extensive searches for optical jets have been carried out (e.g. Butcher \& Miley 1980). As the number of confirmed detections has increased, particularly from the Hubble Space Telescope (HST), some patterns in their properties have emerged. Optical jets show a strong similarity to the radio morphology and a high degree of linear polarisation (e.g. 3C 66B, 3C 78, 3C 264; Perlman et al 2006), but optical jets are shorter and narrower than their radio counterparts (Sparks et al 1995).

NGC 7385 is a low redshift (z=0.026, Wegner et al 1999) elliptical galaxy located in a poor cluster (Zwicky 2247.3+1107) and is also part of the 3CRR sample of radio sources published by Laing, Riley \& Longair (1983). It was mapped with the Westerbork Synthesis Radio Telescope by Schillizi \& Ekers (1975), who found it to have head-tail structure with the tail extending about 510 kpc. The radio luminosity and morphology classify it as Fanaroff-Riley type 1 (Fanaroff \& Riley 1974). It was later observed at higher resolution with the Very Large Array (VLA) by Leahy et al (1999) where a counter jet and an unresolved core could be distinguished. An optical image from the Palomar telescope showed a knot in the counter-jet region, and this was interpreted by Simkin \& Ekers (1978) as a signature of the jet interacting with a relatively dense intergalactic cloud. They detected faint emission lines of OI, OII and H$\beta$ within the knot.

We observed using the Wide Field Camera 3 on the Hubble Space Telescope (HST) in conjunction with \textit{Chandra} as part of a survey to look at the X-ray activity of low redshift 3CRR galaxies. The present paper reports the discovery of an optical and X-ray jet in those data, and gives an improved description of the morphology of the Simkin and Ekers knot.

Throughout this paper we adopt the cosmological parameters $H_0$ = 70 km s$^{-1}$ Mpc$^{-1}$, $\Omega_{\Lambda 0}$ = 0.7, $\Omega_{m 0}$ = 0.3. The redshift of NGC 7385 corresponds to a luminosity distance of 106 Mpc and a projected linear scale of 0.49 kpc/arcsec.

\section{Observations and Analysis}

\subsection{VLA observations}

NGC 7385 was observed in the L band with the VLA as part of programs AL419 and AH129. Table 1 gives the observational details. For these observations, 3C 286 was the primary flux-density and polarisation calibrator. The data reduction and calibration were carried out using the Common Astronomy Software Applications package ({\sc casa}). The calibration was straightforward, closely following the procedures set out in the {\sc casa} cookbook and reference manual.

\begin{table*}
 \centering
 \begin{minipage}{140mm}
  \caption{Observation details for VLA archival data}
  \begin{tabular}{@{}llrrrrlrlr@{}}
  \hline
  Project & VLA & Central & Bandwidth & Integration & Date & Beam size& Noise\\ 
ID & Configuration & Frequencies (GHz) & (MHz) & time (seconds) & & (arcsec) & (mJy/beam) \\
 \hline
AH129 & A & 1.455, 1.515 & 50 & 1760 & 1983 October 1 & 1.6 & 0.7 \\ 
AL419 & C & 1.405, 1.475 & 25 & 11460 & 1997 August 27 & 3.0 & 1.7 \\ [1ex] 
\hline
\end{tabular}
\end{minipage}
\end{table*}

The resulting images of the radio structure of NGC 7385 for the AL419 and the AH129 projects are shown in Figure 1 and Figure 2, respectively, and discussed in Section 3.

\begin{figure}
  \includegraphics[width=8cm]{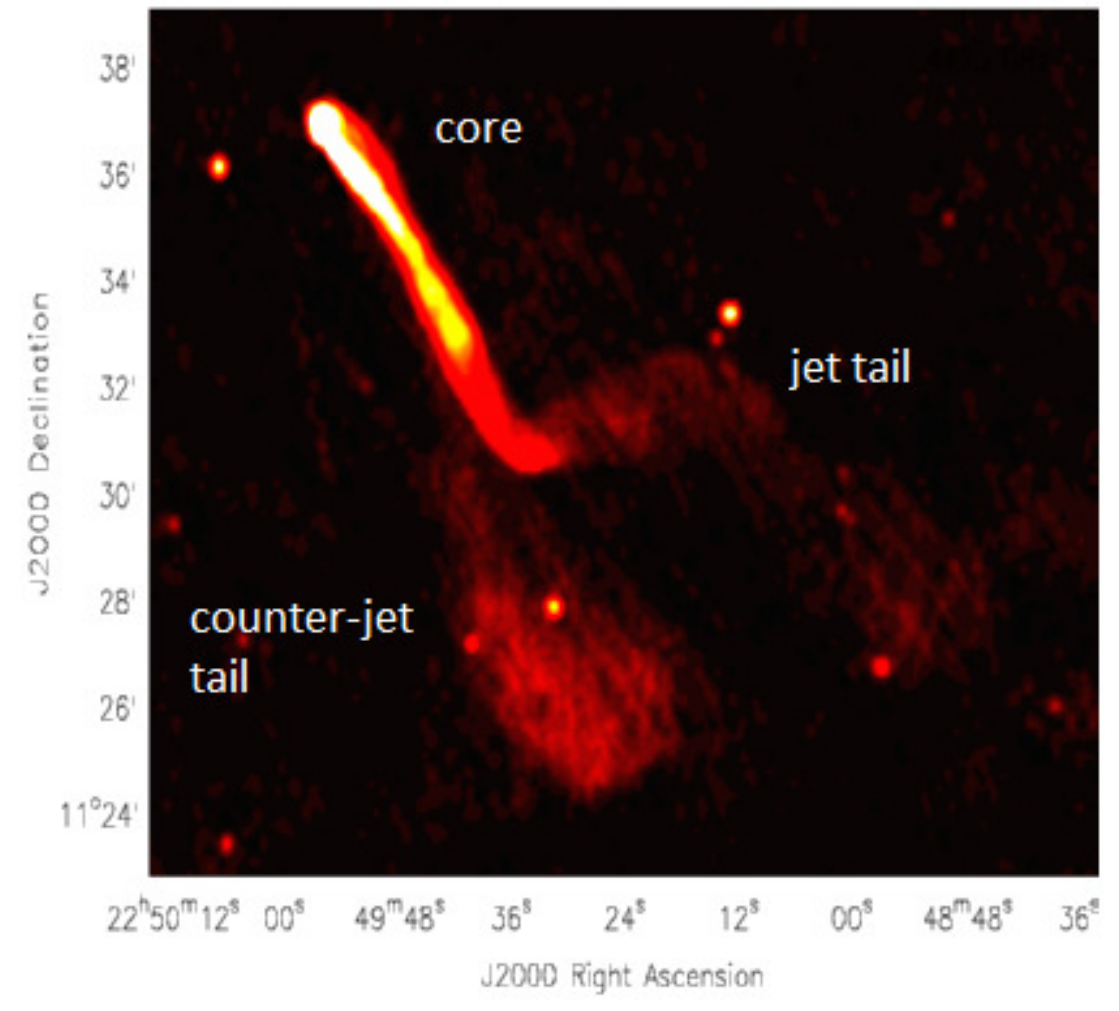}\\
  \caption{Radio image from the archival AL419 project at 1.47 GHz showing the extended radio structure of NGC 7385. One tail appears to have bent to the W and then S. The other is to the S.  }
\end{figure}

\begin{figure*}
  \includegraphics[width=17cm]{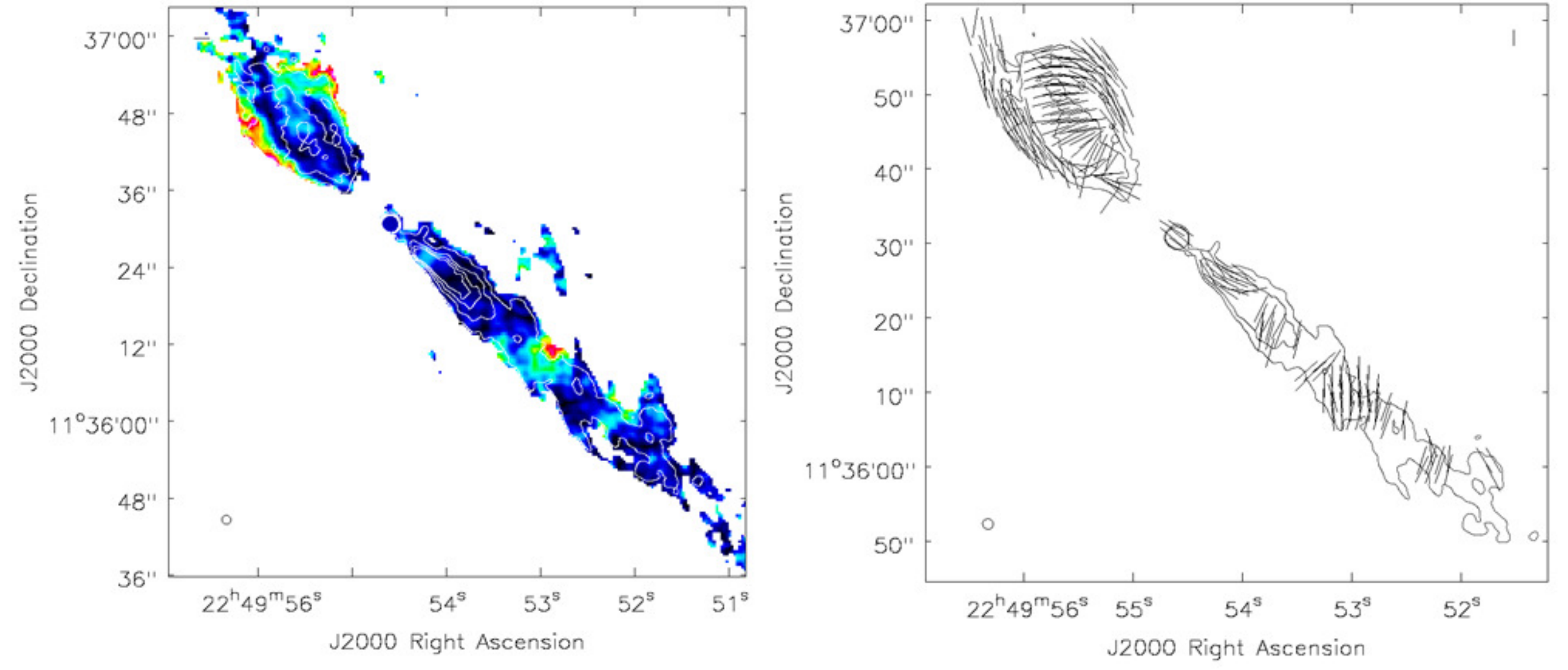}\\
  \caption{\textit{Left}: Fractional polarisation and intensity contours from the AH129 dataset showing the inner jet and counter jet. Contour levels 2, 4, 6, 8 mJy beam$^{-1}$. \textit{Right}: Polarisation rotated by 90\degree  to show the  apparent magnetic field direction. Intensity contours at 2 mJy and 4 mJy beam$^{-1}$ are also shown. }
\end{figure*}

\subsection{HST observations}

HST images of NGC 7385 were obtained on 2009 October 14 with the WFC3 instrument. The observation details are given in Table 2. The data were retrieved from the Multimission Archive at Space Telescope and all the images were flat-fielded, bias-corrected and wavelength calibrated with the standard HST pipeline routines.

\begin{table*}
 \centering
 \begin{minipage}{140mm}
  \caption{Observation details for HST data}
  \begin{tabular}{@{}llrrrrlrlr@{}}
  \hline
Band/Filter & Exposure time (seconds) & Apertures & Central wavelength (angstroms) & Pixel size (arcsec) \\
 \hline
F160W & 2606 & IR-FIX & 15369 & 0.128 \\ 
F775W & 1600 & UVIS & 7647 & 0.0396 \\ 
F555W & 712 & UVIS & 5308 & 0.0396 \\ [1ex] 
\hline
\end{tabular}
\end{minipage}
\end{table*}

The IRAF task \textit{ellipse} was used to fit elliptical isophotes to the galaxy images. A residual image was produced by subtracting an elliptical model from the raw image, thus removing most stellar emission. Good photometry is strongly dependent on accurate knowledge of the sky value. We used sky regions clear of other sources, and took an average to determine the background for each image. The signal was then converted into Jansky, using the pipeline calibration. The flux profile along the jet was extracted and is discussed in more detail in section 3. The same region was used for the radio, optical and X-ray extractions.

The raw image of NGC 7385 with the F160W filter is shown in Figure 3 with the residual image produced after the subtraction of the underlying stellar emission and the removal of cosmic rays. In addition, Figure 4 shows an enlarged version of the optical jet.

\begin{figure*}
  \includegraphics[width=15cm]{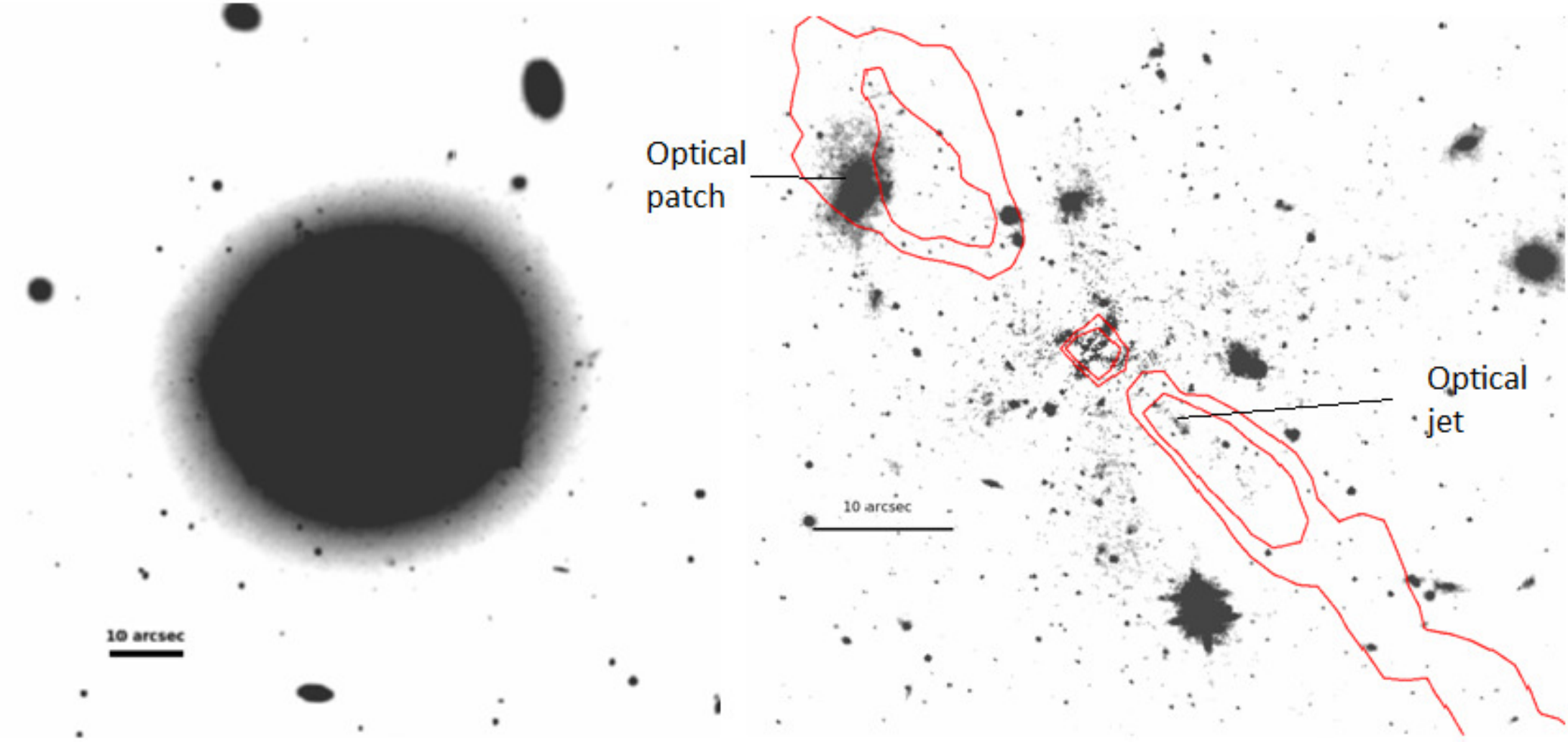}\\
  \caption{HST optical images in F160W. \textit{Left}: before galaxy subtraction. \textit{Right}: Zoomed view after removing cosmic rays and galaxy subtraction. Key optical features are labelled and the AH129 radio image is indicated by contours at 2 mJy and 4 mJy beam$^{-1}$. Both images are displayed on a logarithmic intensity scale. }
\end{figure*}

\begin{figure}
  \includegraphics[width=8cm]{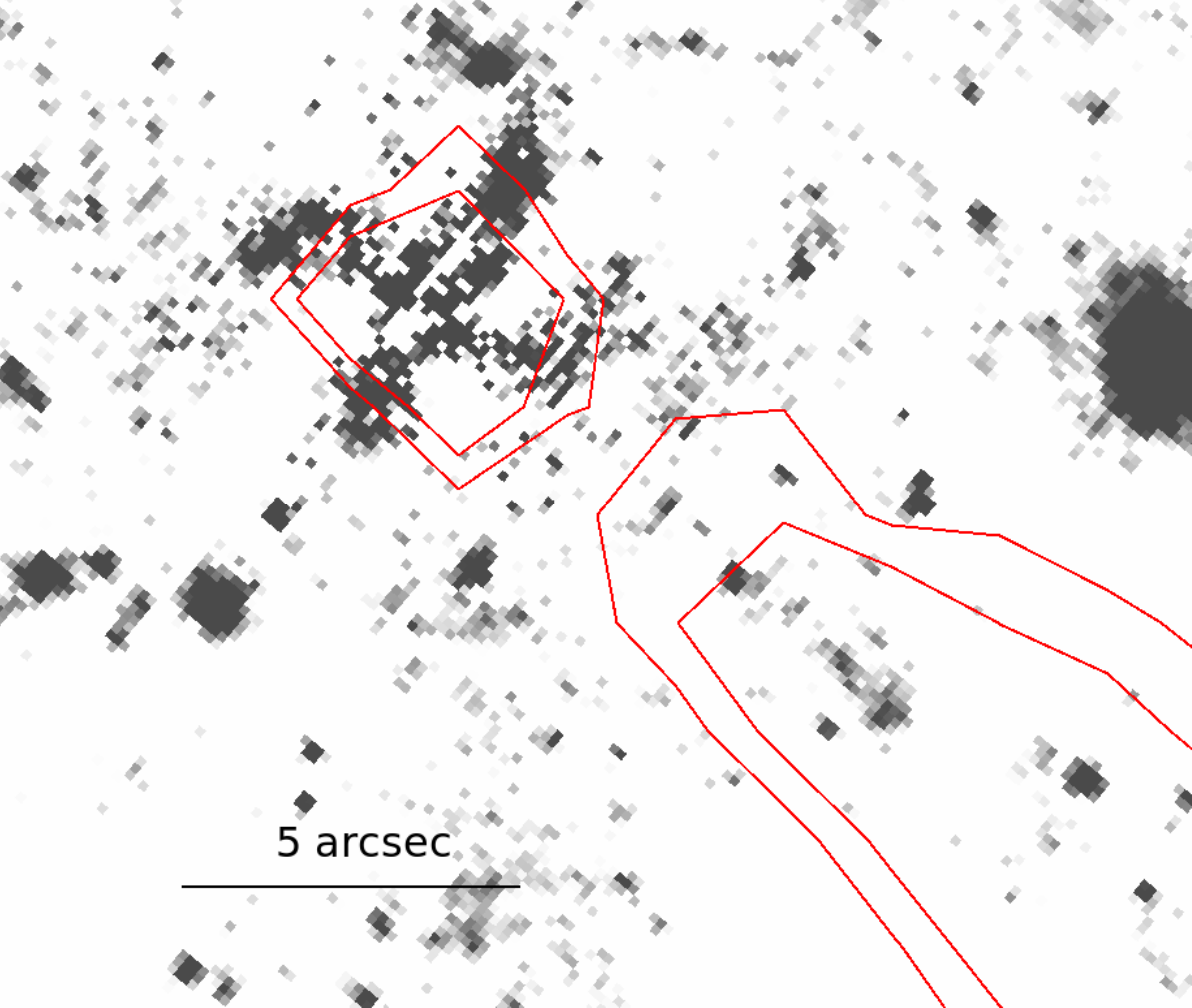}\\
  \caption{Close up of the optical jet from HST data and shown in Figure 3. The radio image is indicated by contours at 2 mJy and 4 mJy beam$^{-1}$}
\end{figure}

\subsection{\textit{Chandra} observations}

The ACIS-I instrument on \textit{Chandra} was used to observe NGC 7385 on 2009 August 15 in \textit{VFAINT} imaging mode for 40 ks, with observation ID 10233. The data were imported into CIAO and reprocessed by following the `science threads' in CIAO version 4.6 to generate a new level 2 events file.  The X-ray image in 0.5-2.0 keV is shown in Figure 5 with the AH129 radio intensity contours overlayed. The left image has been smoothed with a Gaussian function of $\sigma$=3 pixels, where 1 pixel = 0.492 arcseconds, and the right image is unsmoothed.

23 counts were found over the region of the optical jet in the 0.5-2.0 keV energy band, where the counts expected using the average over other regions symmetrically-placed relative to the X-ray core were 4. This gives a probability of detection in excess of 99\%. 

The X-ray flux was extracted using CIAO task \textit{srcflux}, which determines the net count rate and flux for sources in a \textit{Chandra} observation given source and background regions and a spectral model. The background region was an annulus centered on the nucleus of the AGN with inner radius 3$''$ and an outer radius 9.3$''$. The X-ray jet and the counts in the read-out direction were excluded from the background region. This shape ensures the removal of nuclear X-ray emission spread over the jet by the point response. 

\begin{figure*}
  \includegraphics[width=15cm]{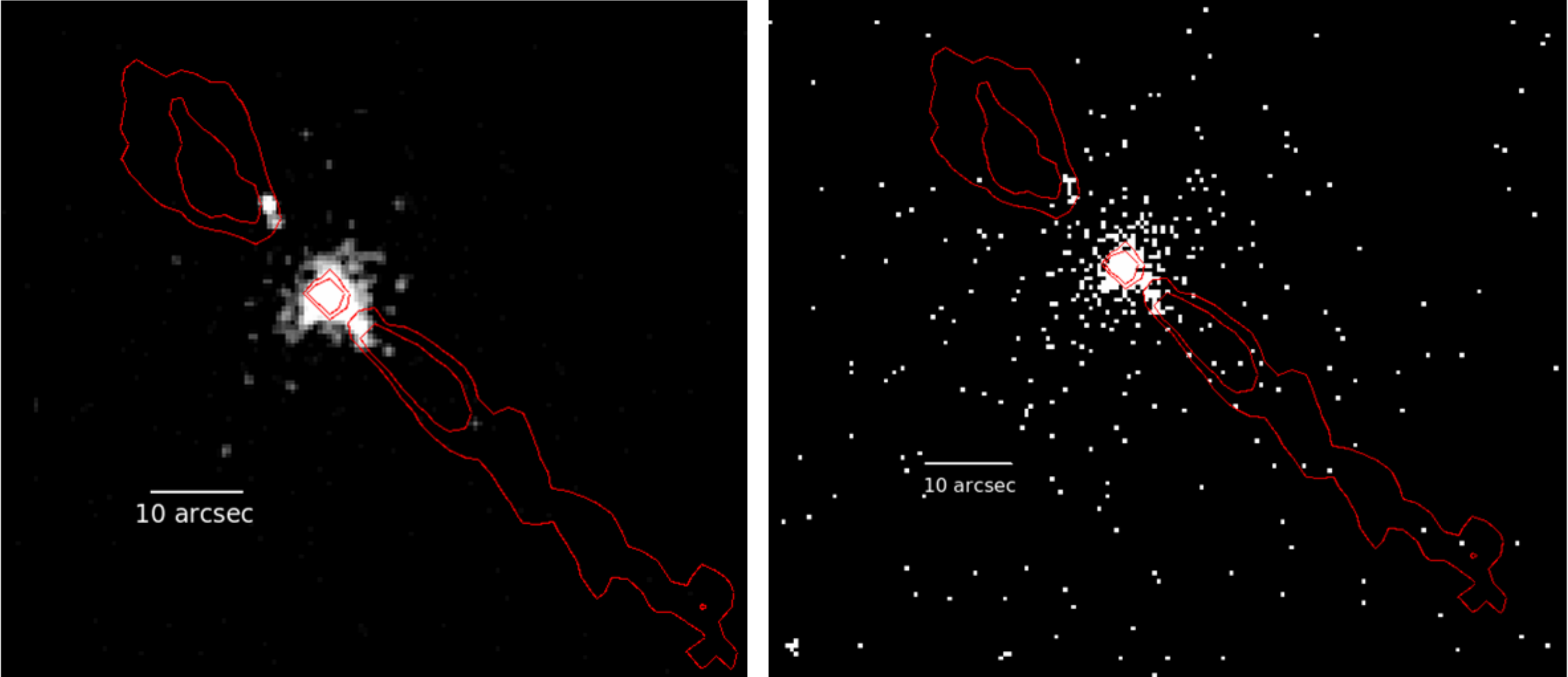}\\
  \caption{The Xray emission of NGC 7385 produced using \textit{Chandra} data and processed in CIAO. The contours are radio intensity contours from the AH129 project with contour levels 2 mJy and 4 mJy beam$^{-1}$. The X-ray jet can be seen extending to the south west away from the bright core. \textit{Left}: The image has been smoothed with a Gaussian function $\sigma$=1.5 arcsec, and the energy range is 500-2000 eV.  \textit{Right}: Unsmoothed image showing the raw counts in the source with energy range 0.5-2 keV.}
\end{figure*}

\section{Results}

\subsection{Jet profiles}

\subsubsection{Radio emission}

The VLA radio images show a one-sided radio jet extending to the SW of an unresolved core. The NE counter-jet is not detected close to the nucleus, but expands into a lobe-like feature. The full length of the jet is evident from Figure 1 and the collimated part of the emission extends about 7 arcminutes (200 kpc). The image shows extended emission in which two separate tails appear SW of the core at roughly 300 kpc from the nucleus. The straight SW jet becomes diffuse and bends to the W to enter the westernmost tail. The second tail is the end of the counter-jet, produced through a series of bends and a 180\degree turn back towards the SW about 30$''$ (15 kpc) from the core (Schilizzi \& Ekers 1975).

The structure of the NE component is clear from Figure 2. It can be interpreted as a stalled jet, and its shape suggests that the whole structure is being affected by an interaction with the external medium. The counter-jet turns to the NW and then to the SW as evident (Figure 1) by a broadening of the radio structure. The stalled jet then crosses the main jet and emerges S of the nucleus as a plume. 

Figure 2 shows the fractional polarisation, $P$, of the jet and the magnetic field direction. The fractional polarisation in the counter jet appears to increase towards the edges up to a maximum of $P$ = 0.65. In the centre of the structure the polarisation is as low as 0.05. The apparent magnetic field is perpendicular to the jet axis in the low polarisation region, and rotates to become parallel at the edges where they appear to track the edges of the source.

The unresolved core has low fractional polarisation, $P$=0.1, and the magnetic field lies parallel to the jet direction. Further down the jet, the polarisation direction starts to rotate, and about 14$''$ from the core they become perpendicular. Beyond this region the fractional polarisation increases steadily up to maximum values of 0.6 at 30$''$ before decreasing again. In this region of strong polarisation the vectors are at 50\degree  to the jet axis. There is another region of high polarisation around 46$''$ from the core and here the field vectors are again perpendicular to the jet. Figure 6 shows the variation of fractional polarisation with distance along the centre of the jet. There is a clear region of low polarisation at 13-14$''$, corresponding to where the vectors rotate, and therefore likely caused by beam depolarisation.

\begin{figure}
  \includegraphics[width=8cm]{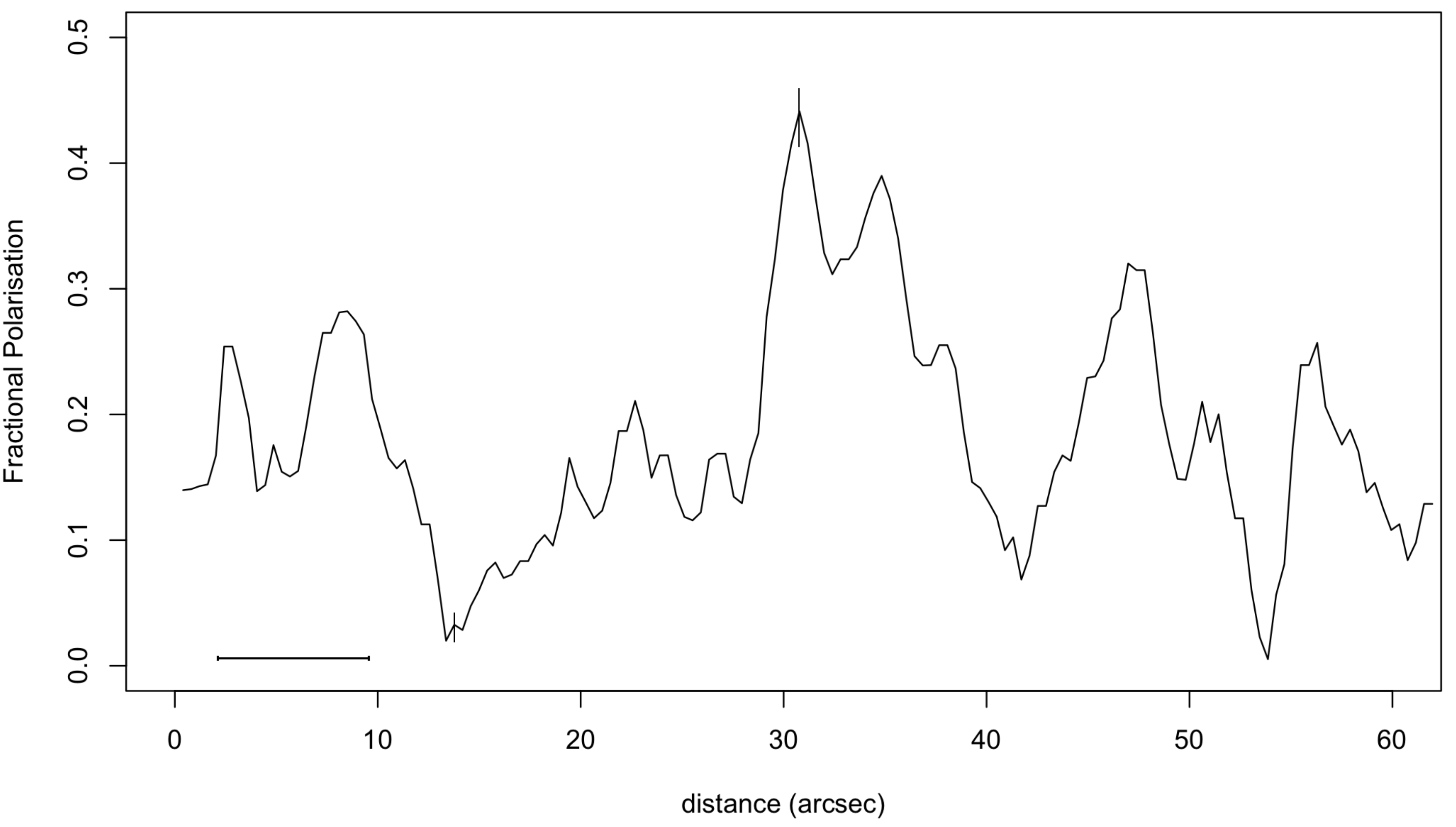}\\
  \caption{The change in fractional polarisation with distance along the centre of the radio jet. The projection used was 1.6$''$ wide at an angle 319\degree   from the nucleus. The bar marks the bright part of the optical jet.}
\end{figure}

\subsubsection{Optical emission}

The optical jet can be seen in Figures 3 and 4 extending to the SW within the radio contours. Close to the nucleus the subtraction is imperfect so we can only examine the structure and spectral distribution from 3$''$ onwards. The intrinsic faintness of the jet makes the edges difficult to define. The jet is brightest and best defined between 3-10$''$ and so this region was used as the area over which to extract the flux, although extended optical emission can be seen up to 17$''$ from the core. Errors were evaluated by combining the Poisson error in the counts on the jet with the uncertainty due to subtracting the galaxy. A noise term due to the background was also included. Although we recognise that optical emission in the region of the jet could be due to background galaxies, it is highly unlikely that such emission would exhibit a linear pattern coincident with the centre-line of the radio jet of NGC 7385.

\subsubsection{X-ray emission}

The X-ray jet is brightest in a region about 4-6$''$ from the core (Figure 5). Fainter emission is detected over a further 5$''$. It is difficult to determine the significance of the faint X-ray emission beyond the inner jet, however the expected background in the region of the outer jet is 3 countsl. In this region, we detect 28 counts in 15 pixels along 2$''$ of the jet, and since the Poisson probabilty of detecting these counts when 3 are expected is much less than 0.1\%, this implies a significant detection.

There is also a second region of X-ray emission along the W contours of the counter jet. This has no optical counterpart in the HST images. There are 26 counts within this bright X-ray structure, more than for the jet. This X-ray emitting region has been listed as CXO J224944+113640 in the \textit{Chandra} source catalogue (Evans et al 2010).

\subsubsection{Surface brightness profiles}

Multi-frequency surface brightness profiles of the jet are given in Figure 7 up to a distance of 16$''$ (8 kpc) from the core. Regions have been labelled A-F in accordance with the position of peaks in intensity in the optical jet.  A profile at 90\degree to the jet was also examined toconfirm the flatness of the background, and to assess the errors associated with the surface brightness along the jet. The width of the profiles (1.1$''$) was chosen in accordance with the dimensions of the optical jet, so that off-axis stars would not be included.

\begin{figure}
  \includegraphics[width=8cm]{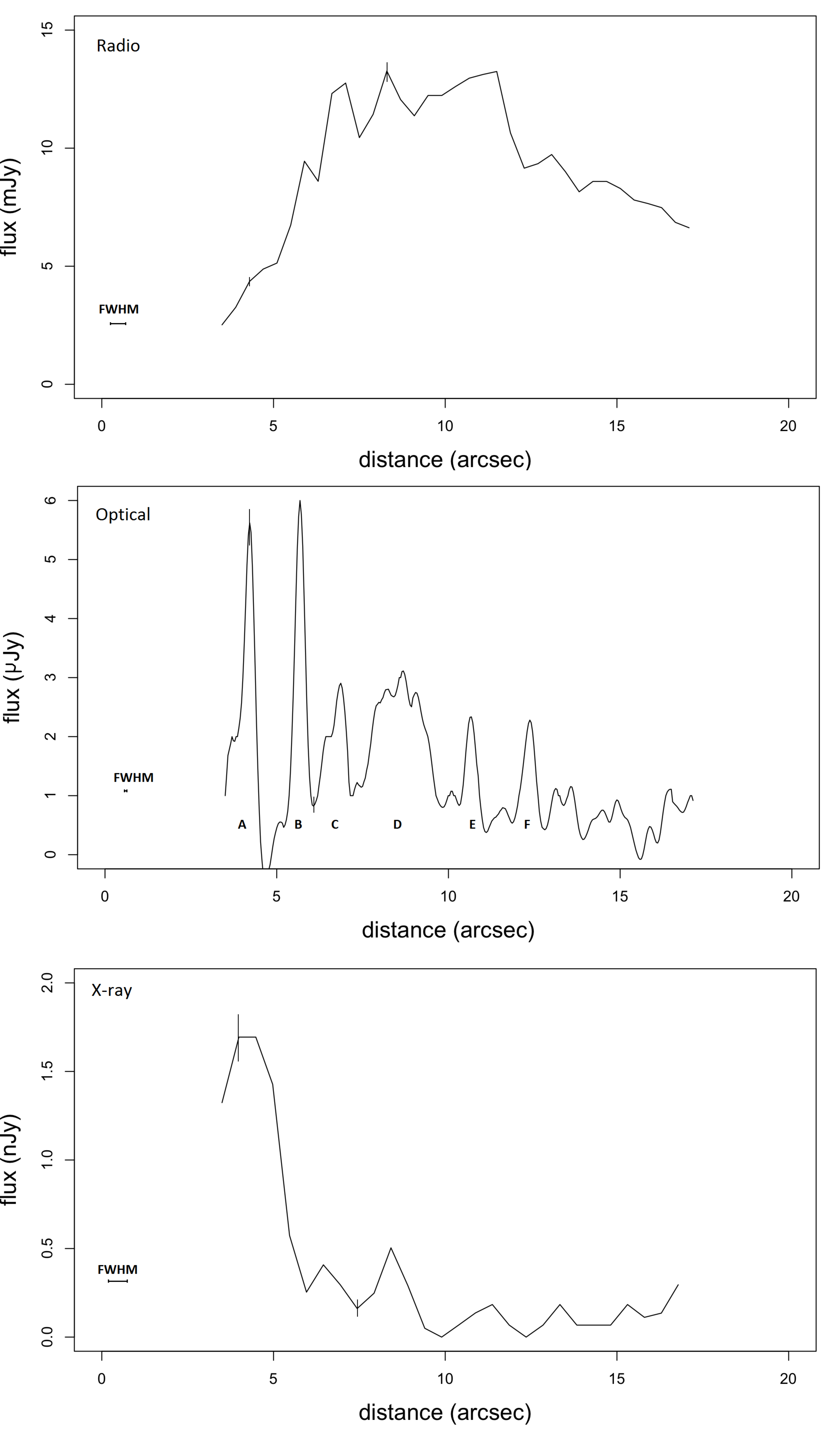}\\
  \caption{Surface brightness profiles along the centre of the radio, optical and X-ray jets. The same projection was used for each frequency, 1.1$''$ wide out to distance 16$''$ from the core. 6 peaks in the optical profile have been labelled as knot A-F. It was not possible to extract profiles closer to the core than 3.5$''$ due to imperfections from the ellipse subtraction. All data were background subtracted and the beam size is shown for each frequency. Representative error bars are also included.}
\end{figure}

The profiles begin at 3.5$''$ from the centre of the core and extend out to 16$''$. In regions closer to the core in the optical image, galaxy and nuclear subtraction are uncertain with bad residuals. A spectral energy distribution was extracted for regions A-D, which are bright at all frequencies, with results in all 3 bands converted into Janskys using default calibration. 

The optical and X-ray plots show a peak at knot A. In the radio this feature is not prominent, possibly indicating a compact knot of 1 mJy at 1.51 GHz. The optical peak region B corresponds to a decrease in X-ray intensity and a gradual intensity increase in the radio. Surface brightness was measured at each pixel along the length of the defined jet region and the high resolution of the HST image means that regions A and B can be distinguished. Due to the lower resolution of \textit{Chandra}, it is possible that the large X-ray peak arises from separate unresolved components. Region C shows a peak at each frequency, as does region D. The positions of the peaks in region E and F in the X-ray image are approximately 1$''$ further along the jet than in the optical and none of the peaks are well resolved in the radio.  The X-ray brightness quickly drops off  after region B. The radio intensity increases steadily out to 7$''$ and the jet remains bright beyond the X-ray and optical extracts. Optical knot F coincides with the radio polarisation rotation region where the parallel magnetic field rotates to become perpendicular to the jet direction.

\subsection{Multi-frequency spectral distribution}

The spectral index is defined as $\alpha$ where $S \propto \nu^{-\alpha}$ and the values obtained by extracting the flux (S) from the images are given in Table 3. 

The flux in the X-ray jet was extracted using \textit{srcflux} in CIAO. The data were analysed for uncertainties in the source and the background, and the flux at 1 keV was calculated. An adopted X-ray power law of $\alpha_X$ = 1.1 (energy index) and a Galactic absorption of $N_{H}$ = 5.06 $\times$ 10$^{20}$ cm$^2$ obtained from the COLDEN program, gave a 1 keV jet flux density of 10.6 $\pm$ 0.7 nJy.

\begin{table}
 \centering

  \caption{Spectral indices for each frequency band}
  \begin{tabular}{@{}llrrrrlrlr@{}}
  \hline
Region & Spectral index \\ 
 \hline
$\alpha_{r}$ & 1.1$\pm$ 0.65 \\ 
$\alpha_{ro}$ & 0.48$\pm$ 0.21 \\ 
$\alpha_{o}$ & 1.1 $\pm$ 0.32 \\ 
$\alpha_{ox}$ & 1.25 $\pm$ 0.15 \\ 
$\Delta\alpha$ & 0.77 $\pm$ 0.36 \\ [1ex] 
\hline
\end{tabular}

\end{table}

Spectral indices both within and between frequency bands are given in Table 3. To compute the radio-to-optical spectral index, $\alpha_{ro}$, both radio frequencies and the lowest optical frequency were used. For the optical-to-X-ray index, $\alpha_{ox}$, the highest optical frequency was used.

The spectral range of our data is too small to give a useful radio spectral index, $\alpha_{r}$, so we adopt $\alpha_{r}$=0.5, as in Schilizzi and Ekers (1975). They plotted the spectral index of emission along the tail of NGC 7385 with distance out to 14 arcminutes at 0.408 GHz and 1.415 GHz. They found the nucleus to have a flat spectrum, with $\alpha_r$ = 0.5, and the tail component becoming progressively steeper with distance, reaching $\alpha_r$ = 1. Robertson (1981) also found $\alpha$ = 0.5 along the first 4 arcminutes of the radio jet, using WSRT data at 1.4 GHz and 5.0 GHz. A similar spectral index in the NE radio component was found in the S tail, consistent with the idea that it is a counter-jet ejected in the direction of motion of the galaxy.

There are no published results for the flux densities in the first 9$''$ of the jet from the nucleus at any frequency. Our radio index has a large uncertainty due to the small frequency range used to extract the flux density. Matched resolution multi-frequency radio observations of NGC 7385 would be very useful for constraining the spectral index. In the optical too, the range spans only 0.3 decades in frequency, giving a small pivot in the spectral-index fits.

It is difficult to find strong trends in observed features of multi-frequency radio galaxies similar in luminosity to NGC 7385 since relatively few have been detected in both optical and X-ray.

For all optical jets, the $\alpha_{o}$ is in the range of 1.2-1.4, similar to our value for NGC 7385. If the optical slopes were steeper, the jet emission would be more difficult to detect since they would in general be fainter and so jets with a steeper optical spectral index tend to be selected against. However flatter-spectrum jets would not be selected against, so the absence of flat optical slopes could be significant.

In M87, Biretta et al (1991) published jet spectra showing a general trend for the optical spectral index to steepen with distance from the core. We would need deeper optical data to test if this is the case in NGC 7385.

\begin{figure}
  \includegraphics[width=8cm]{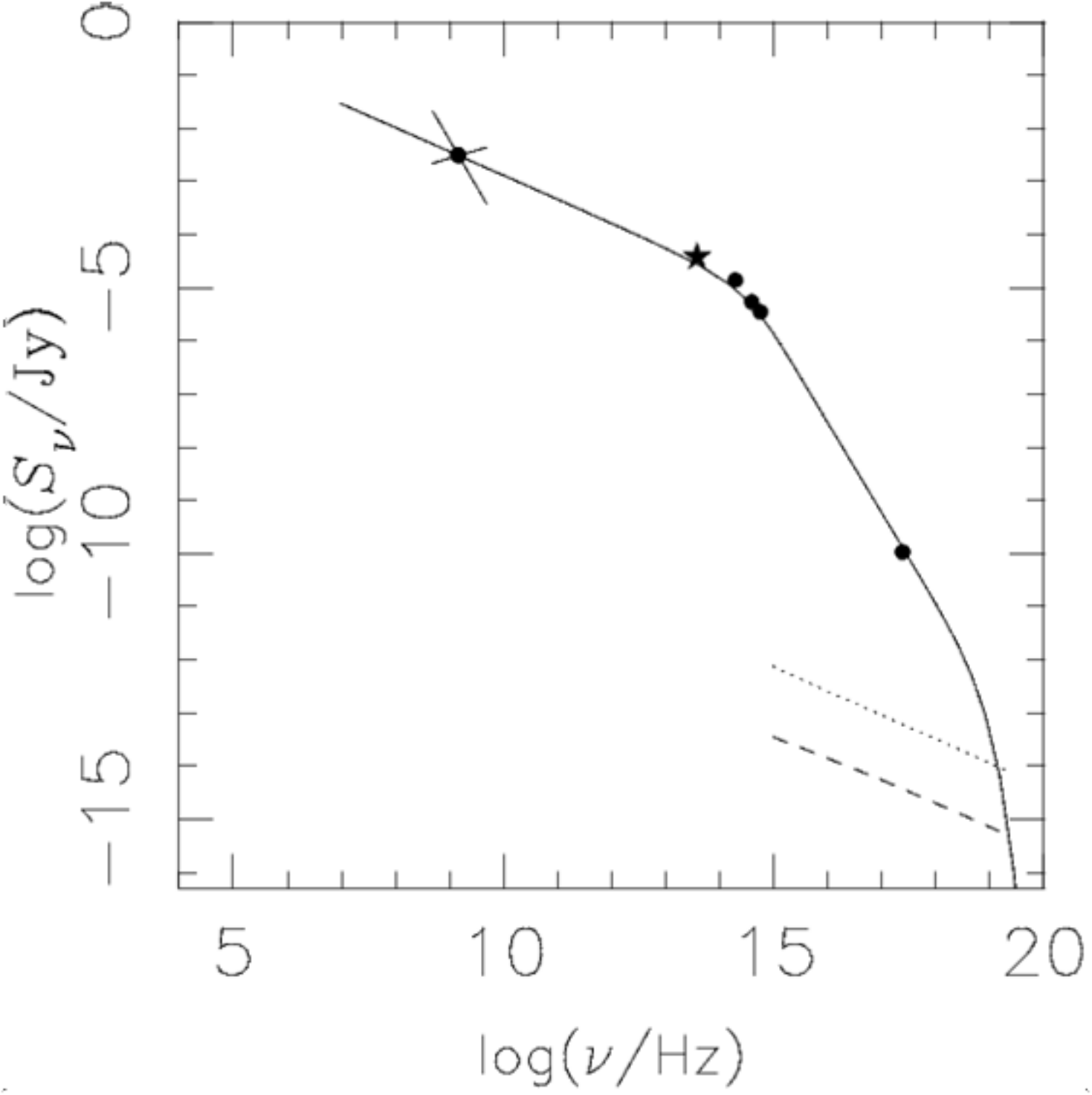}\\
  \caption{The spectral distribution of the jet in NGC 7385. The shape is consistent with synchrotron radiation (solid black line). The contribution to the X-ray emission from synchrotron self Compton (dashed line) and inverse Compton scattering of the CMB (dotted line) is also shown. There is a spectral break of 0.75 in the optical-infrared region and the bowtie shows the error range in the radio spectral index. The black star shows the 8.0 $\mu$m flux from Spitzer data}
\end{figure}

\subsection{Model fitting}

A spectral fit based on the method of Hardcastle, Birkinshaw \& Worrall (1998) is shown in Figure 8. We fitted a synchrotron spectrum and found the magnetic field strength, based on the assumption that the electrons and magnetic field are in equipartition and that relativistic beaming is negligible. The method uses the derived magnetic field strength to calculate the contribution to the X-ray emission from the synchrotron self Compton process and inverse Compton scattering of the cosmic microwave background radiation (CMB), both of which are found to be negligble. The spectrum shows the need to introduce a break $\Delta\alpha\sim$ 0.8 into the synchrotron spectrum. The break is constrained by the X-ray flux density and needs to be at $\sim$ 9 $\times$ 10$^{13}$ Hz, in the infrared-optical region. Our derived equipartition magnetic field strength of 2.3 nT is strongly dependent on the minimum energy of the electrons, which was set at 5.11 MeV, i.e. with Lorentz factor $\gamma_{min}$ = 10. With the aim of constraining the spectrum further, we extracted the flux density from a Spitzer image of NGC 7385 (see Figure 8) with wavelength 8.0 $\mu$m and found that the results are consistent with a near-IR position for the break frequency.

The knots at E and F are at 10.6 and 12.5$''$ from the nucleus respectively, corresponding to distances of 5 and 6 kpc. The synchrotron lifetime for emission is given by
$t_{cool} = 0.043  B^{-\frac{3}{2}} \nu^{-\frac{1}{2}}$   in years, for an electron emitting in a B field (Tesla) with frequency $\nu$ in Hertz (Worrall \& Birkinshaw 2006). For optically emitting electrons at 6.45 $\times$ 10$^{14}$ Hz, and using the equipartition magnetic field strength 2.3 nT, this gives a synchrotron lifetime of the order 10$^4$ yrs. This is smaller than the light travel time from the nucleus, but only by a factor of two. However if we assume that the X-ray emission is also produced via the synchrotron process, the lifetime for electrons at knot D is around 800 yrs. This is over 10 times less than the light travel time to this region. This means that at some point the particles must have been accelerated or reaccelerated within the jet. The lifetime for electrons producing radiation at the break frequency is of the same order as the light travel time to the end of the optical jet. The need for the reacceleration of high energy electrons to the outer regions of jets is common in many other similar sources. The sites of reacceleration are often identifiable by the location of shocks.

\subsection{Other optical features}

There is an additional optical feature located within the radio counter-jet lobe. This bright optical knot has been described by Simkin \& Ekers (1979). They detected faint OII, OIII and H$\beta$ emission lines with the Palomar telescope and suggested that the knot is not an optical counterpart of the process responsible for the radio emission but arises from a secondary interaction with a relatively dense intercluster medium. The systematic velocity of the gas relative to the stars in the galaxy strongly suggests that the knot is interacting with the radio jet. Their ground-based images from the Palomar Observatory in June 1975 show no evidence of the optical jet.

In the galaxy-subtracted HST image in Figure 3 the knot is bright and appears to extend beyond the radio contours. The position of the radio flux maximum as indicated by the contours is separated from the centre of this optical knot by approximately 3$''$, and it is also notable that the radio contours are distorted at the structure. A three-colour composite image was created using the F160W filter as the red image, F775W as green and F555W as blue.  The optical knot was found to be significantly bluer than the rest of the galaxy, suggesting it is gas rich and a site of strong star formation. The current radio data are not adequate to search for Faraday rotation in the knot, and therefore to discover whether the structure is in front or behind the radio counter-jet. For this we would need new radio data.

A surface brightness profile of the elliptical fit to the galaxy in the HST data shows a disordered pattern close to the core. This suggested that a dust lane may be concealed close to the nucleus. The dust lane becomes apparent in a three colour composite of the raw HST images (Figure 9). The dust lane is roughly perpendicular to the direction of the optical jet and on the counter-jet side suggesting a disk and jet tilted so that the optical jet is moving towards us. The measured position angles of the dust lane and optical jet are 236\degree $\pm$ 5 and 317\degree $\pm$ 2 respectively. Dust features are observed in many radio-loud elliptical galaxies surrounding the AGN; Madrid et al (2006) found nine galaxies out of a sample of 69 low redshift galaxies to have dust lanes (e.g. 3C 285, 3C 321, 3C 430). Capetti et al (2000) found the fraction of host galaxies with dust lanes to be higher at low radio power, with 54\% of FRI sources in the sample to contain dust lanes. NGC 7385 should be added to statistical studies of the dust content of low power radio galaxies.

\begin{figure}
  \includegraphics[width=8cm]{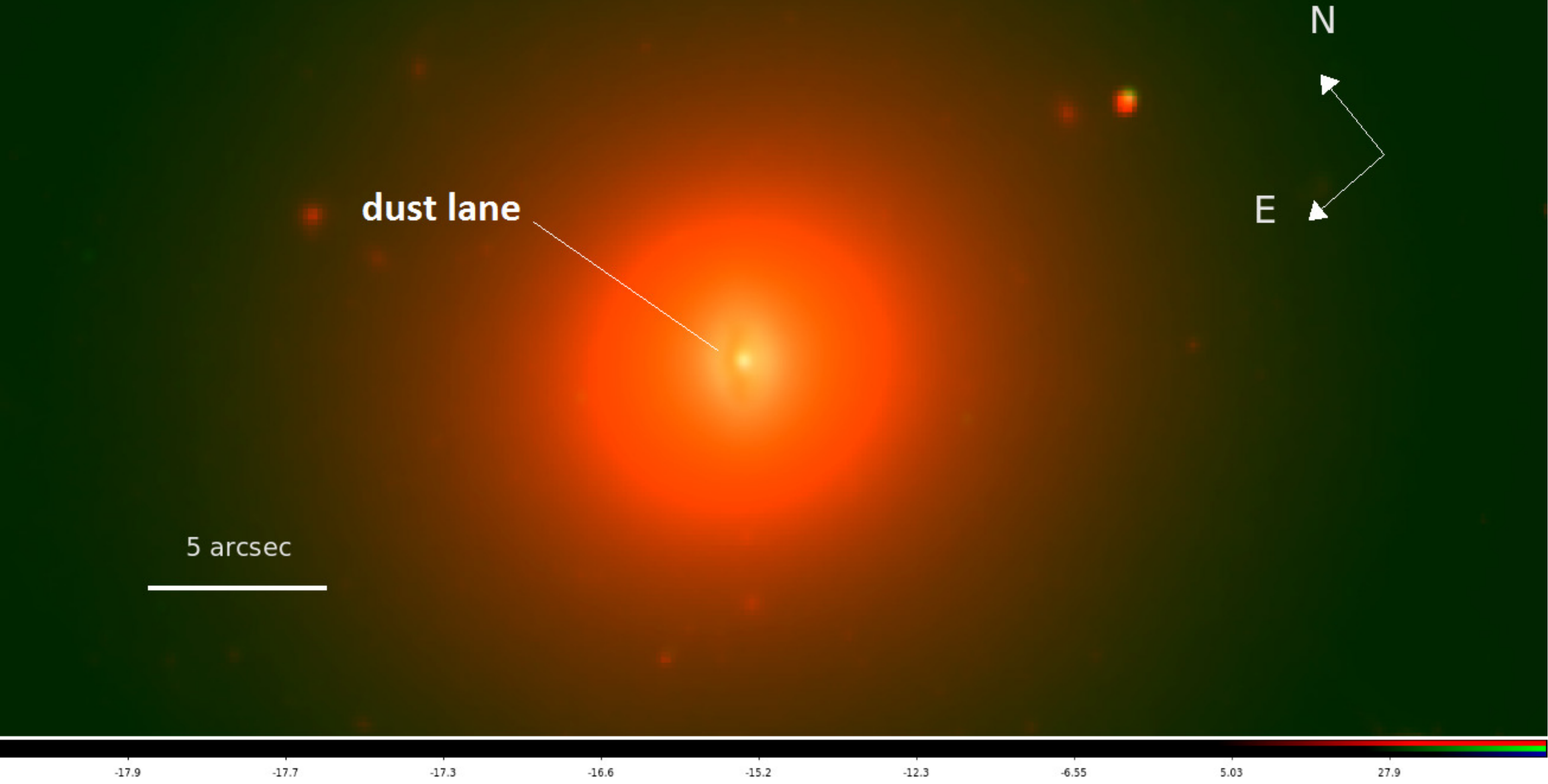}\\
  \caption{The dust lane close to the nucleus on the counter-jet side in a three colour composite image. The F160W filter was defined as red, F775W as green and F555W as blue. }
\end{figure}

\section{Discussion}

In the preceding sections we have discussed the polarisation and magnetic field structure in the radio jet of NGC 7385, and presented new images of an optical jet obtained through HST imaging. The spectral indices for the jet from the radio through to the X-ray are presented in Table 3 and Figure 8 and we find the synchrotron mechanism to be responsible for the observed emission in all bands. In this section we discuss the physical processes at work within the jet, and compare this to other known jets with radio and optical polarimetry. In particular we look at the relationship of the jets in M 87 (Perlman et al 1999), 3C 66B, 3C 264 (Perlman et al 2010) and 3C 15 (Dulwich et al 2007) to this source. 

\subsection{Polarisation and magnetic field structure}

Because of the narrow frequency range in these data we cannot map the rotation measure near NGC 7385, and so cannot infer the source's magnetic field structure with any confidence. In addition, the centre line of the jet appears to have a low polarisation channel and the magnetic field vectors are disordered close to the core. This is likely the result of beam depolarisation, because the resolution of the radio image (1.6$''$) is poorer than the angular scale over which the source is coherent.

Nevertheless, the polarisation data are helpful in some regions. In the counter-jet we find areas of high linear polarisation near the edges, where the magnetic field appears to follow the flux contours. This suggests shearing of field lines along the edge of the counter-jet lobe. This is particularly prominent at the northern end of the counter-jet where the field appears to track the jet bends, suggesting a strong link between the local magnetic field and the dynamics that bend the jet. In the centre of the component,  compression of the magnetic field would cause perpendicular field lines and this magnetic field structure has been observed in many other radio lobes (Hogb{\H o}m 1979).

An increase in fractional polarisation near the edges of jets could also be evidence of torsional shear. This polarisation increase is seen 14$''$ from the nucleus along the northern edge of the main jet. This would not be so easily observed in the centre of the jet because of the cancellation of different field directions seen in projection. Knot A in the jet of M 87 has been modelled as a torsional shock (Bicknell \& Begelman 1996, Perlman et al 1999), and similar features are seen in Cen A (Clarke et al 1992). 

The structure in the magnetic field can also be explained in terms of velocity gradients along the jet. A sheath-like parallel field structure is shown in both the counter jet and the jet. This suggests that there could be two distinct regimes in the jet modeled by a fast moving relativistic core or `spine', surrounded by a shear layer (the `sheath') decelerating due to entrainment with external material. This spine-sheath model would predict the magnetic field structure observed in the counter-jet, where the slow outer sheath surrounds a core which is faint due to beaming away from the observer.

Laing (1996) uses a spine-sheath model to explain the rotation of the magnetic field vectors from parallel to a jet axis to perpendicular further out. This change in vector direction is suggested in Figure 6 at 13$''$ by a region of depolarisation. This field structure has also been observed in many other FRI jets (e.g. 3C 66B, Hardcastle 1996). Laing explains the pattern in terms of shear layers with a longitudinal component and a jet core without. Close to the nucleus the core is highly relativistic and beamed out of the direction of the observer, so that the parallel magnetic field is seen by the observer. The jet decelerates as it moves away from the nucleus and the perpendicular component of the field becomes more important, causing the change in the field direction.  In this model, emission from both longitudinal and transverse B fields can only be seen if it is beamed towards us. The radio polarisation vectors in 3C 15 (Dulwich et al 2007) are consistent with this model, and it is suggested that the magnetic field is ordered with a helical structure. Determining whether the polarisation structure observed is due to a helical magnetic field structure can prove difficult. A key piece of evidence for the presence of helical fields is the detection of rotation measure gradients across the jet. Higher resolution radio data for NGC 7385 could therefore constrain the dynamics of the jet flow.

\subsection{Emission mechanisms}

We have focused on modeling the spectral distribution using synchrotron radiation. Contributions to the X-ray emission from inverse Compton scattering of the CMB and the synchrotron self Compton process are predicted to be negligible (Figure 8). A broken power law electron energy distribution fits the data well, with an electron power law of energy slope $2\alpha + 1$ =1.9 with a break at energy $ 9 \times 10^{11}$ eV. The equipartion magnetic field strength, B$_{eq}$= 2.3 nT, is somewhat lower than other measurements for sources with optical jets (e.g. 33 nT for 3C 264; Perlman et al 2010). The minimum energy, or $\gamma_{min}$, is the uncertain factor with the greatest influence on the value of the magnetic field strength. B$_{eq}$ is roughly proportional to $\gamma_{min}^{-(2\alpha -1)/(\alpha + 3)}$ for $\alpha$ $>$ 0.5 (Worrall \& Birkinshaw 2006). A smaller value of $\gamma_{min}$ therefore would lead to an increase in the magnetic field strength. However we have already assumed a low value of $\gamma_{min}$ = 10, so we could not recover a field as large as in 3C 264. The volume of the jet is also important, and it was modelled using the width of the optical jet. 

The lifetime of electrons emitting optical and X-ray synchrotron radiation are shorter than the light travel time for electrons to the outer knots. Mechanisms which could permit such a situation include a highly relativistic flow, a `low loss channel' for transporting electrons, and local reacceleration (Hardcastle 1996). In low loss channels electrons might be transported in a region of low magnetic field to the outer regions of jets, and adiabatic and synchrotron losses might be negligible compared to losses due to inverse Compton scattering of the CMB photons. However in the X-ray, inverse Compton scattering on the CMB and starlight would further limit the lifetime of electrons making this mechanism an unlikely explanation for the observed emission.

Many other jets exhibit a similar energy loss problem and local reacceleration is widely believed to resolve the problem. Looking at the spectral indices of the knots individually would constrain possible sites for particle acceleration, since the optical indices would be distinguishable from interknot regions as significant losses between knots would be observed. Hardening of indices in knots close to the nucleus could suggest that the knots are shocks at which first order Fermi acceleration is taking place (Hardcastle 1996). Alternatively, particles could be accelerated by a second order Fermi process driven by turbulence through the jet (Bicknell 1994). This process however would produce a shear sufficient to maintain a parallel magnetic field direction throughout the jet (Leahy 1991) and this is not observed in NGC 7385.

Offsets between peaks at different frequencies could constrain the location of these privileged sites of particle acceleration. For the jet in 3C 66B, Hardcastle et al (2001) found an offset between the radio and X-ray peaks and suggested a model in which radio knots are the sites of particle acceleration with a shock close to the X-ray peak. The shift in the position of the peak would be caused by rapid synchrotron losses of the particles emitting in the X-ray as they propagate away from the shock. Figure 7 shows an offset in the position of the optical and X-ray peaks at knots E and F, so the physical process for NGC 7385 could be similar to that described above. In addition, the low polarisation region 12$''$-13$''$ from the core, and coincident with the position of the optical flux maximum in knot F, shows a rotation in the magnetic field from parallel to the jet axis to perpendicular. Perlman et al (2010) describe this same pattern for the structure of the knot A/B region in 3C 264, and suggest that the low polarisation could be the result of field superposition and that the angle change may be due to shocks at the location of the maxima. Alternatively, the angle change in knot C in 3C 15 was modeled as the superposition of a helical strand of magnetic field on a more ordered field structure (Dulwich et al 2007). Unfortunately we do not have optical polarimetry data to help distinguish between these two models for NGC 7385.

Theoretical and observational work from Brunetti et al (2003) showed that low-power radio hot spots are optimum candidates for detecting an optical counterpart. They found that the shape of the synchrotron spectrum in these regions is strongly influenced by the magnetic field strength, constraining the position of the break frequency. They produced a plot of the synchrotron break frequency as a function of the equipartition magnetic field for data from the VLT and also from literature (see their Figure 4). The plot showed a clear trend which suggested that large values of  break frequency are likely due to the effect of the magnetic field on the cooling of the electrons. Our calculated value of 2.3 nT predicts a value of break frequency close to 10$^{15}$ Hz, as inferred (Figure 8), although in our case we see this in a jet knot rather than a hot spot, so relativistic effects may be present. 

The five jets in M 87, 3C 78, 3C 371, 3C 66B and 3C 264 show the general shape of the spectrum in Figure 8. For a simple energy-loss model of synchrotron radiation, the spectral break would be 0.5. We measured the break to be 0.77 (but with a large error), in accordance with the jets of other low power radio galaxies where the break is in the range 0.6-0.9 (e.g. 3C 78, 3C 371, 3C 66B; Worrall 2009). Explanations for this steeper break have been offered (Pacholczyk 1970, Wilson 1975, Reynolds 2009) in terms of a decrease in the magnetic field away from sites of particle acceleration, with the assumption that electrons propagate towards regions with a lower magnetic field. Another plausible explanation uses the idea of velocity gradients in the jet flow (Section 4.1). Georganopoulos \& Kazanas (2003) propose that for small angles to the line of sight, the spectral break should be 0.5 as suggested by theory. However at higher angles, when the emission of the spine is being beamed out of our line of sight, spectral breaks can be steeper. The degree of alignment affects the observed jet emission at high energies, where the optical and X-ray emission should be dominated by the spine for highly aligned jets. The orientation of the NGC 7385 jet is unknown.

While HST imaging observations now exist for almost all known optical jets, HST polarimetry does not. Polarimetry across multiple frequencies can uncover differences in the magnetic field geometry encountered by different radiation populations. With this information, it is then possible to form a three dimensional picture of the energetic and field structure in the jet. Optical polarisation maps have been produced for fewer than 10 jets, and as a result very little is known about the energetic structures. It is not therefore known for the majority of sources to what extent the radio and optical emitting electron populations are cospatial and encounter a common magnetic environment. 

\subsection{The interaction of the optical knot with the counter jet}

Most explanations for optical features observed in the outer regions of radio galaxies invoke an interaction between radio emitting material and gas. Many optical knots have been observed to have optical emission lines and an underlying optical continuum (Simkin \& Bicknell 1984). These interactions can form stars (DeYoung 1981) or produce shock excited gas (Graham \& Price 1981). The optical emission can be explained by stellar emission from newly formed stars, or alternatively by emission from the plasma. For the knot in NGC 7385, emission lines have been detected (Simkin \& Ekers 1978). Simkin \& Bicknell suggest that the emission-line ratios could be explained either by shock heating or photoioinisation by a power-law continuum, as would be the case if a cloud was entrained by the radio counter-jet. The blue continuum emission from the knot supports the idea that emission is from newly formed stars. Interactions of optical gas clouds and radio jets can provide velocity measurements not obtainable through radio observations alone. However such interactions are rare, and only a few such sources have been studied in depth. 

PKS 2152-699 is a bright radio galaxy observed to host a high-ionisation cloud near a radio knot associated with a jet lobe (Worrall et al. 2012). There is evidence for a strong cloud-jet interaction, in which the radio jet appears to be deflected due to oblique shocks within the cloud. Multiwavelength data of that galaxy have helped to reveal the location of jet bends and ionisation properties of the gas cloud.Unlike PKS 2152-699 however, NGC 7385 has no X-ray emission in the region of the optical knot, showing the absence of hot-phase gas coexisting with optical emission-line gas. The jet deflection in NGC 7385 is more extreme than in PKS 2152-699, because the jet is disrupted and forms a radio lobe, destroying the collimation of the counter-jet. It is not currently possible to investigate the geometry of the interaction of the gas cloud with the radio emitting plasma because the archival VLA polarisation data is relatively poor. This prevents a detailed analysis of rotation measure gradients in this area which would help locate the cloud with respect to the jet. More detailed optical spectroscopy of the cloud would be useful to measure its kinematics.

\section{Conclusions}

We report the discovery of optical and X-ray jets in the radio galaxy NGC 7385. This radio source also displays prominent jet bends and an interaction with an optical gas cloud. 

The shape of the radio to X-ray electron spectrum along the jet is consistent with synchrotron radiation, with a low magnetic field strength $\sim$ 2 nT. NGC 7385 can now be placed on the short list of radio sources with synchrotron jets detected in both the optical and X-ray, with the hope that more extensive investigations into the optical jet structure could reveal the sites for particle reacceleration.

Archival VLA data revealed the structure of the radio polarisation in the counter-jet and along the main jet. Regions of high polarisation along the counter-jet edges suggest a strong link between the magnetic field and the dynamics that bend the counter-jet. The presence of a bright, optical knot in close proximity to the counter-jet provides a unique opportunity to investigate strong cloud-jet interactions. With new radio observations, a rotation measure map could help constrain the position of the optical cloud with respect to the radio structure and further explore the magnetic field structure and jet energetics within these regions. 

\section*{Acknowledgments}

JR acknowledges funding from STFC. We thank the referee for helpful suggestions to improve the manuscript. The National Radio Astronomy Observatory Very Large Array is operated by Associated Universities Inc., under cooperative agreement with the National Science Foundation. This research has used observations made with the NASA/ESA \textit{Hubble Space Telescope}, obtained from the data archive. We thank the CXC for its support of \textit{Chandra} observations, calibrations and data processing.

\section*{References}

\noindent Baade W., 1956, ApJ, 123, 550 \\
\noindent Bicknell G. V., 1994, ApJ, 422, 542 \\
\noindent Bicknell G. V., Begelman M. C., 1996, ApJ, 467, 597 \\
\noindent Biretta J. A., Stern C. E., 1991, AJ, 101, 1632 \\
\noindent Biretta J. A., Sparks W. B., Macchetto, F., 1999, ApJ, \\
\indent 520, 612 \\
\noindent Birkinshaw M, Worrall D, M., Hardcastle M, J., 2002, \\
\indent MNRAS, 335, 142 \\
\noindent Bohringer H. et al., 2001, A\&A, 365, 181 \\
\noindent Bridle A. H., Perley R. A., 1984, ARA\&A, 22, 319 \\
\noindent Bridle A. H., 1996, in Hardee P. E., Bridle A. H.,\\
\indent Zensus J. A., eds, ASP Conf. Ser. 100, Energy \\
\indent Transport in Radio Galaxies and Quasars, \\
\indent San Fransisco, 383, \\
\noindent Brunetti G., Mack K., Prieto M., Varano S., 2003, \\
\indent MNRAS, 345, 40 \\
\noindent Butcher H. R, Miley G. K., 1980, ApJ, 235, 749 \\
\noindent Capetti A., de Ruiter H. R., Fanti R., 2000, A\&A, 362, 871 \\
\noindent Clarke D, A., Bridle A, H., Burns J. O., Perley R, A., \\
\indent 1992, ApJ, 385, 173 \\
\noindent De Young D. S., 1981, Nature, 293, 43 \\
\noindent Dulwich F., Worrall D. M., Birkinshaw M., Padgett \\
\indent C. A., Perlman E. S., 2009, MNRAS, 398, 1207 \\
\noindent Dulwich F., Worrall D. M., Birkinshaw M., Padgett \\
\indent C. A., Perlman E. S., 2007, MNRAS, 374, 1216 \\
\noindent Evans I. N. et al., 2010, ApJS, 189, 37 \\
\noindent Fanaroff B. L., Riley J. M., 1974, MNRAS, 167, 31 \\
\noindent Georganopoulos M., Kazanas D., 2003, ApJ, 594, 27 \\
\noindent Ghisellini G., Tavecchio F., Chiaberge M., 2006, \\
\indent MNRAS, 368, L15 \\
\noindent Graham J. A., Price R. M., 1981, ApJ, 247, 813 \\
\noindent Hardcastle M, J., 1996, MNRAS, 278, 273 \\
\noindent Hardcastle M, J., Birkinshaw M., Worrall D, M., 1998, \\
\indent MNRAS, 294, 615 \\
\noindent Hardcastle M, J., Worrall D, M., Birkinshaw M., 2001, \\
\indent  MNRAS, 326 1499 \\
\noindent Hjorth J., 1995, ApJ, 425, 17 \\
\noindent H{\H o}gbom J, A., 1979, A\&A, 36, 173 \\
\noindent Jester S., Harris D. E., 2006, ApJ, 648, 900 \\
\noindent Laing R. A., Riley J. M.,  Longair M, S., 1983, \\
\indent MNRAS, 204, 151 \\
\noindent Laing R. A., 1996, in Hardee P. E., Bridle A. H.,\\
\indent Zensus J. A., eds, ASP Conf. Ser. 100, Energy \\
\indent Transport in Radio Galaxies and Quasars, \\
\indent San Fransisco, 241 \\
\noindent Leahy J, P., 1991, in Hughes P. A., ed., Beams and Jets \\
\indent in Astrophysics. Cambridge: Cambridge Univ. Press, \\
\indent p100\\
\noindent Leahy J, P., 1993, In: [Roser \& Meisenheimer 1993] \\
\noindent Madrid J. P. et al., 2006, ApJs, 164, 307 \\
\noindent O'Dea C, P., 1999, AJ, 117, 1143 \\
\noindent Pacholczyk A. G., 1970, Radio Astrophysics. San Francisco, \\
\indent CA: Freeman \\
\noindent Perlman, E. S., et al., 2010, ApJ, 708, 171 \\ 
\noindent Perlman E. S. et al., 2006, ApJ, 651, 735 \\
\noindent Perlman E. S., Harris D. E., Biretta J. A., Sparks W. B., \\
\indent Macchetto F. D., 1999, AJ, 117, 2185 \\
\noindent Reynolds S, P., 2009, ApJ, 703, 662 \\
\noindent Robertson J. G., 1981, A\&A, 93, 113 \\
\noindent Roser H, J., Meisenheimer K., 1987, ApJ, 314, 70 \\
\noindent Roser H, J., Meisenheimer K., 1993, Jets in Extragalactic \\
\indent Radio Sources. Berlin: Springer-Verlag \\
\noindent Schilizzi R., Ekers R., 1975, A\&A, 40, 221 \\
\noindent Simkin S. M., Ekers R. D., 1978, AJ, 84, 56 \\
\noindent Simkin S. M., Bicknell G. V., 1984, ApJ, 277, 513 \\
\noindent Sparks W, B. et al., 1995, ApJ, 450, 55 \\
\noindent Wegner G. et al., 1999, MNRAS, 305, 259 \\
\noindent Wetzel R. P., Perlman E. S., 2012, JSARA, 7, 45 \\
\noindent Wilson A. S., 1975, A\&A, 43, 1 \\
\noindent Worrall D, M,. Birkinshaw M., 2006, in Alloin D., \\
\indent Johnson R., Lira P., eds, Lecture Notes in Physics Vol. \\
\indent 693. Springer-Verlag, Berlin, p. 39\\
\noindent Worrall D. M., 2009, A\&AR, 17, 1 \\
\noindent Worrall D. M. et al., 2012, MNRAS, 424, 1346 \\

\end{document}